\documentclass[letterpaper]{article}
\usepackage{iccc}
\usepackage{graphicx}
\usepackage{subcaption}
\usepackage{url}
\usepackage{breakurl}
\usepackage{times}
\usepackage{helvet}
\usepackage{courier}
\title{Is computational creativity flourishing on the dead internet?}
\author{Terence Broad\\
Creative Computing Institute\\
University of the Arts London\\
United Kingdom\\
t.broad@arts.ac.uk\\
}
\begin{document} 
\maketitle
\begin{abstract}
\begin{quote}
The dead internet theory is a conspiracy theory that states that all interactions and posts on social media are no longer being made by real people, but rather by autonomous bots. While the theory is obviously not true, an increasing amount of posts on social media have been made by bots optimised to gain followers and drive engagement on social media platforms. This paper looks at the recent phenomenon of these bots, analysing their behaviour through the lens of computational creativity to investigate the question: is computational creativity flourishing on the dead internet?

\end{quote}
\end{abstract}

\section{Introduction}

The dead internet theory is a conspiracy theory that emerged in the late 2010's or early 2020's that states that large parts of the internet, in particular on social media are no longer occupied by humans and human generated content, but rather posts by AI-driven bots that are designed to control or influence human behaviour \cite{illuminati2021dead}.  

Whist the theory emerges from the fringes of the internet, stemming in conspiratorial thinking as a way of explaining broad-based changes to society from nefarious actors, many commentators have observed that there is a grain of truth to the theory \cite{tiffany2021dead}. With the emergence and widespread adoption of generative deep learning, which is able to generate media in many domains to a plausible human level of fidelity, the possibility of bots acting autonomously or semi-autonomously on social media in order to garner influence is no longer a speculative possibility, but a real phenomenon on many social media platforms. 

Engagement driven social media bots, or \textit{AI influencers} \cite{walter2024artificial}, that are optimised to maximise common social media metrics now exist on many social media platforms such as X (formerly Twitter), Facebook, Instagram, Reddit and TikTok. A recent report investigating web bots and internet traffic estimates that nearly 50\% of web traffic is now driven by bots \cite{imperva2024bad}. 

Bots on social media for the purposes of generating art, poetry or other kinds of content that are designed explicitly as computationally creative agents or artistic experiments in themselves are not new \cite{veale2018twitterbots}. What is new about the phenomenon of these \textit{AI influencers} is that they are not explicitly listed as bots, but instead posing as real people on social media or accounts that aggregate and share the creative work of real people. It is likely these bots are being developed by spammers and scammers to create accounts that have a large amount of engagement so that they traffic off the platforms to content-farm sites where advertising revenue can be generated \cite{koebler2024facebook}.

This paper will investigate the recent emergence of bots on various social media platforms, that appear to be explicitly optimised to drive engagement, and are acting in an at least somewhat autonomous fashion. This paper will analyse the output and behaviour of these models through the lens of computational creativity to determine if these bots are acting as autonomously computationally creative systems in the wild.

\section{Social Media and Content Farms}

Social media platforms, such as Facebook, Instagram, Reddit and TikTok rely on content generated by users as the means of engagement on their platforms and to leverage network effects through an `architecture of participation' \cite{oreilly2005web}. The business model of these platforms is primarily to serve advertisements alongside user generated content. 

Since the invention of the Facebook `News Feed' in 2006, content is usually presented to users on these sites on a homepage, showing them the content from accounts that they follow \cite{arrington2006new}. For many years the presentation of the feed has not been given in a chronological manner, but recommendation algorithms are used to to personalise the experience for users and will prioritise some content over other content in an algorithmic fashion. Since 2022, Facebook's Feed algorithm has been pushing content onto feeds from accounts that users do not follow to compete with TikTok's popular `For You Page' \cite{heath2022facebook} which employs collaborative filtering to personalise the feeds of users to show them content from creators they primarily do not follow \cite{boeker2022empirical}.

Content farms are websites or media creation organisations that make low effort and low quality content that makes money from advertisements, traditional online content farms would have paid freelancers to write articles that would get traffic from organic search, news aggregators and social media traffic \cite{bakker2012aggregation}. With the decline in revenues from organic search and the changes made to search engines and social media platforms that discouraged users to leave their platforms and prioritised rankings for established organisations, content farming has moved to social media platforms themselves in search of revenue through content creator payments, advertisement revenue, and through promoting spam content. 

\section{AI-Powered Content Farming}

Algorithmic content farms have existed long before the recent wave of generative AI tools. Content farms that make animated musical videos on the platform YouTube that take advantage of the young children's passive engagement with YouTube and it's auto-play featured that is powered recommendation algorithms, have been widely observed and criticised \cite{bridle2017something}. The new wave of content farming is making use of generative AI to produce spam content and is being actively promoted by the platforms themselves into unsuspecting users feeds \cite{diresta2024spammers}. These bots accounts appear to be making use primarily of modern text-to-image generation using techniques such as latent diffusion \cite{rombach2022high} and large language models for text generation such as the Generative Pre-Training (GPT) class of models \cite{radford2018improving} to automate text generation on the click-farm websites. 

\begin{figure}[t]
\begin{minipage}[c]{0.49\textwidth}
\label{fig:designs}
\includegraphics[width=\textwidth]{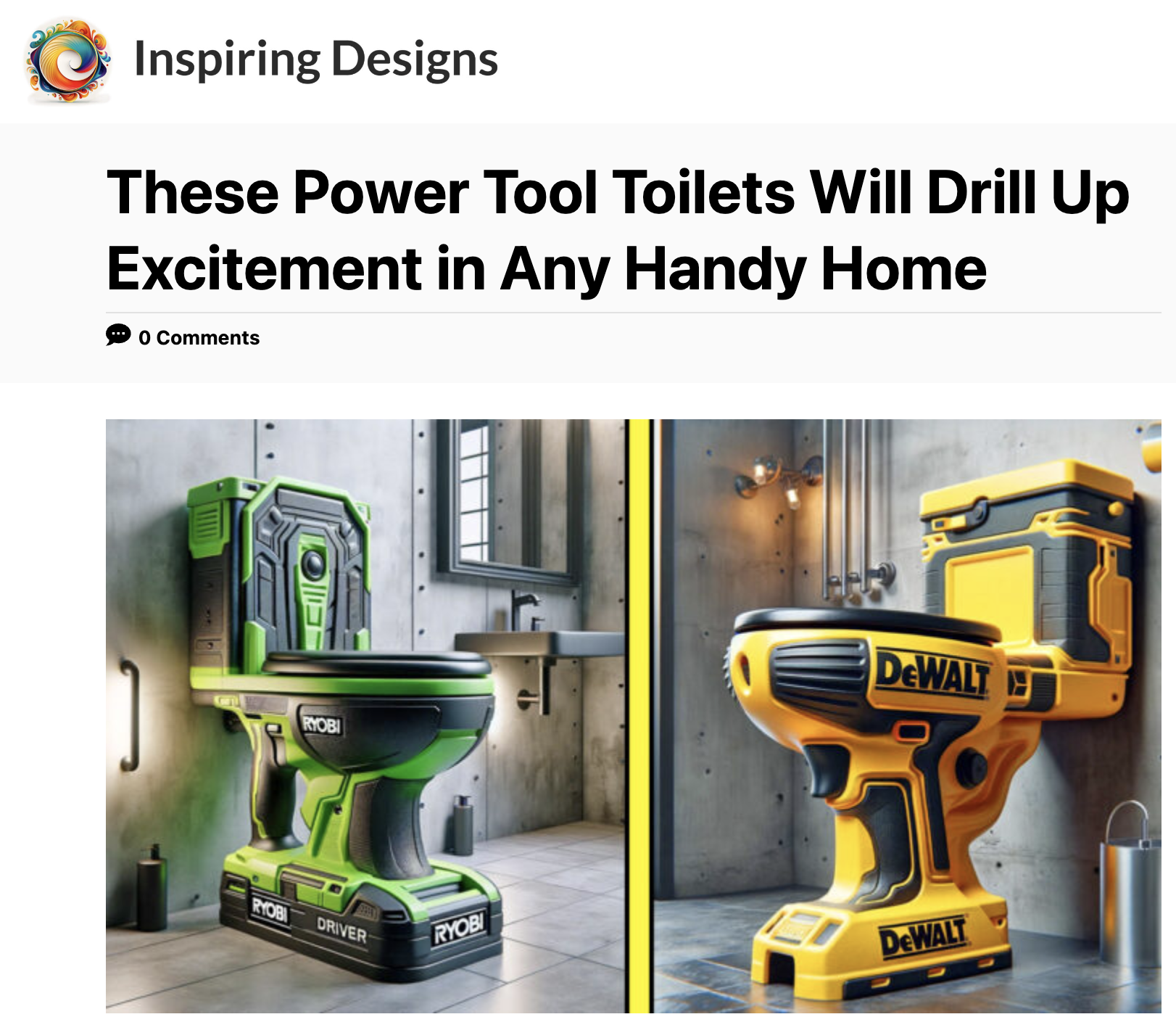}
\end{minipage}\hfill
\caption{Screenshot of the headline of an article from the website Inspiring Designs, detailing the fictional product category of `power-tool toilets'. Available at\: \url{https://inspiringdesigns.net/power-tool-toilets/}}

\end{figure}

An example of an AI-powered content farm is the social media account \textit{Inspiring Designs} and associated website (\url{inspiringdesigns.net}), active on the Facebook, Instagram and TikTok platforms (with hundreds of thousands of followers on each platform). The posts and articles document new fictional product categories, such as `power-tool toilets' (see Figure \ref{fig:designs}) `cowboy-hat showers' and `pickup-truck strollers'. The articles are very likely written by large language models and are written in a highly consistent style and structure. The goal of this site appears to be in order to host ads and generate revenue from affiliate links to the e-commerce website Amazon for products that bear little resemblance to the ones described in the articles. 

Another example of a content farm that makes use of generative AI is the TikTok account \textit{Globetrots}. Globetrots does not fully use generative AI, but rather amalgamates content from satellite imagery, publicly available information based on national and international statistical rankings (i.e. `top 5 most dangerous motorways in the UK', or `top 10 Biggest IKEA stores in the USA'). This account make use of text-to-speech voiceovers which are available to all users in TikTok and widely used by AI spam content producers \cite{koebler2024inside}. These videos are sometimes set to musical melodies and`sung' by AI in various different genres.\footnote{Such as this pop-punk musical countdown of the top 10 worst schools in the UK ranked by Ofsted reports \url{https://vm.tiktok.com/ZGeVG15kd/}.}

\section{`Shrimp Jesus' and Combinatorial Creativity}

\begin{figure}[t]
   % \centering
   \begin{subfigure}[b]{0.225\textwidth}
     \centering
     \includegraphics[width=\textwidth]{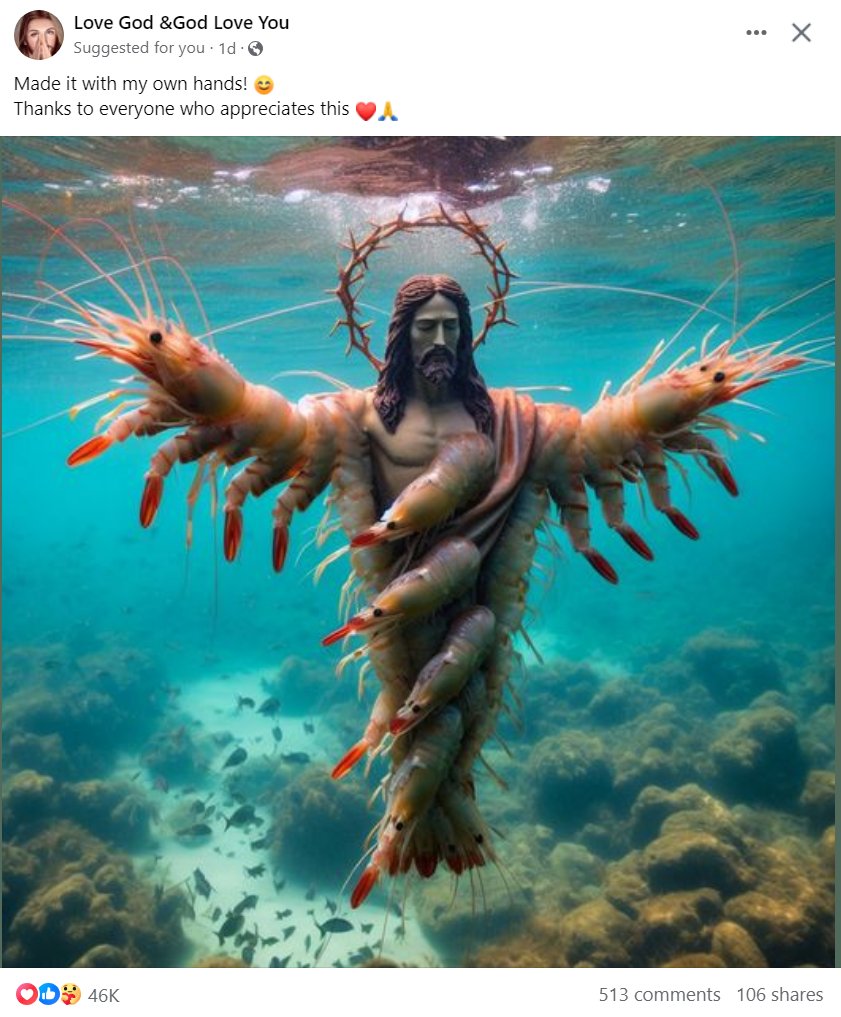}
     \caption{}
     \label{fig:p1}
   \end{subfigure}
   \hfill
   \begin{subfigure}[b]{0.225\textwidth}
     \centering
     \includegraphics[width=\textwidth]{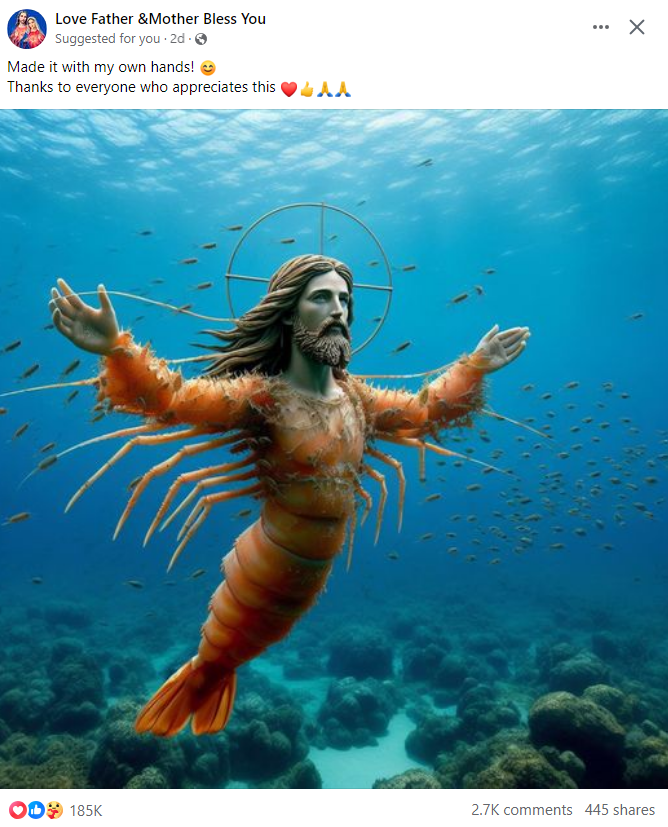}
     \caption{}
     \label{fig:p2}
   \end{subfigure}
  \caption{Images posted on the (now defunct) \textit{Love God \&God Love You} (a) and the \textit{Love Father \&Mother Bless You} (b) Facebook pages. Both posts are AI generated images depicting `Shrimp Jesus' -- the figure of Jesus in part by the anatomical features of Shrimps underwater, posted with the caption \textit{`Made with my own hands! Thanks to everyone who appreciate this'}.} 
  \label{fig:shrimp}
\end{figure}

One of the images that quickly rose to prominence and the attention of this phenomenon on Facebook was the now infamous `Shrimp Jesus' (Figure \ref{fig:shrimp}). This became a widely shared and discussed example of how many accounts on social media were using automated methods to combine concepts to make images and drive engagement on posts. The nonsensical combinations of concepts combined with the mechanical regularity with which images are posted by theses accounts are the main explanatory factors that have convinced many people that pages were being run in a partly or wholly autonomous fashion \cite{koebler2024facebook}.

Many people assume that these images were not just generated by a person entering text prompts into a text-to-image generator, but that some form of automated process was instrumental in creating the text prompts used to generate these images. This is evidenced by the mechanical regularity with which these accounts are posting new images and continuously trying new combinations of visual concepts, regardless of how nonsensical they may be. If these accounts are in fact driven by automated algorithms then,whatever underlying process that is generating these images is clearly an example of combinatorial creativity \cite{boden2004creative} or bisociation \cite{koestler1964act}, where more than one concept is brought together to make a new concept.

\section{Religious Pareidolia}

Fascination with religious pareidolia (the perception of religious iconography in otherwise random of ambiguous patterns) has long existed \cite{obadia2018urban}. Tabloid newspapers are regularly writing articles about people seeing images of Jesus in toast \cite{willis2014second} and various other objects.

\begin{figure}[t]
   % \centering
   \begin{subfigure}[b]{0.225\textwidth}
     \centering
     \includegraphics[width=\textwidth]{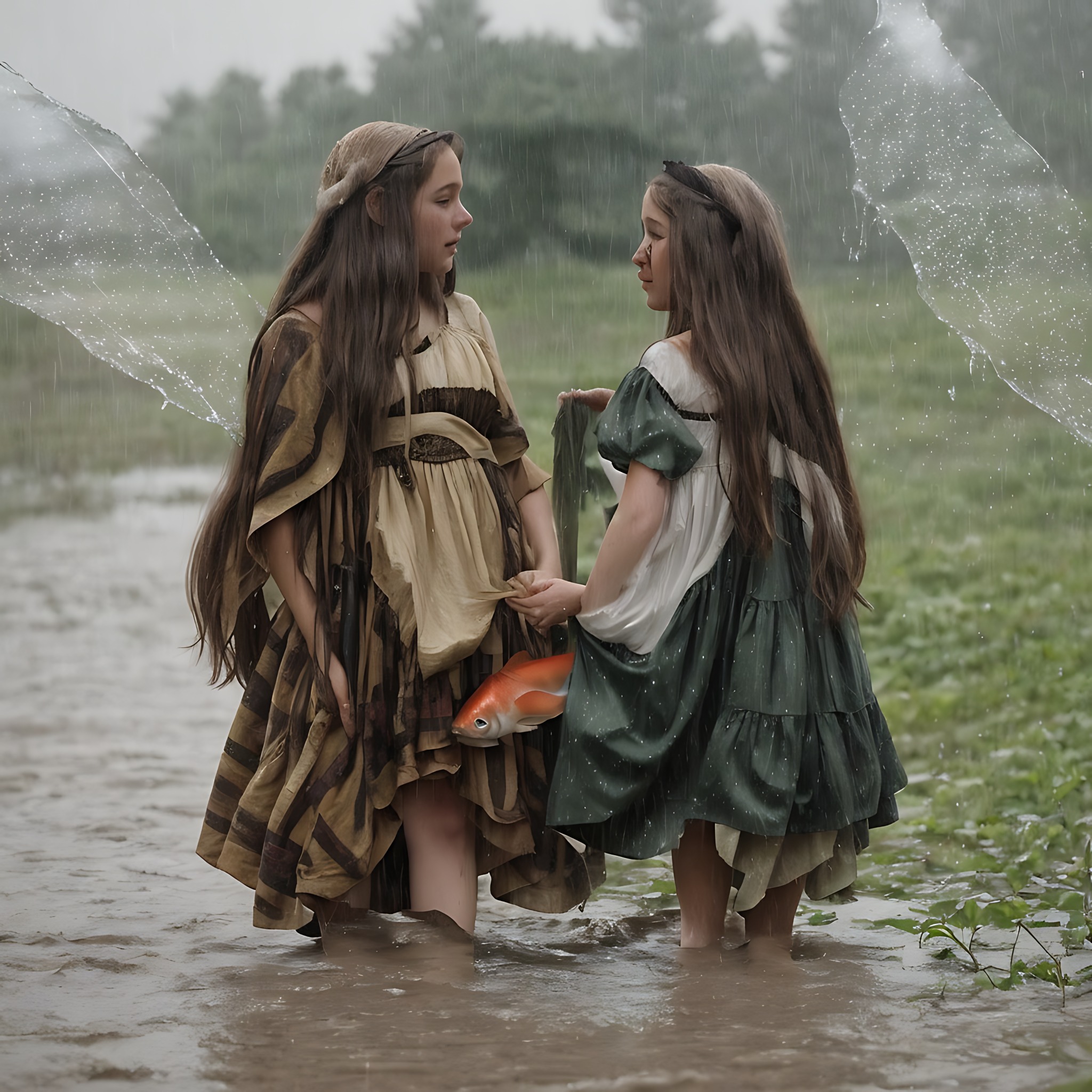}
     \caption{}
     \label{fig:p1}
   \end{subfigure}
   \hfill
   \begin{subfigure}[b]{0.225\textwidth}
     \centering
     \includegraphics[width=\textwidth]{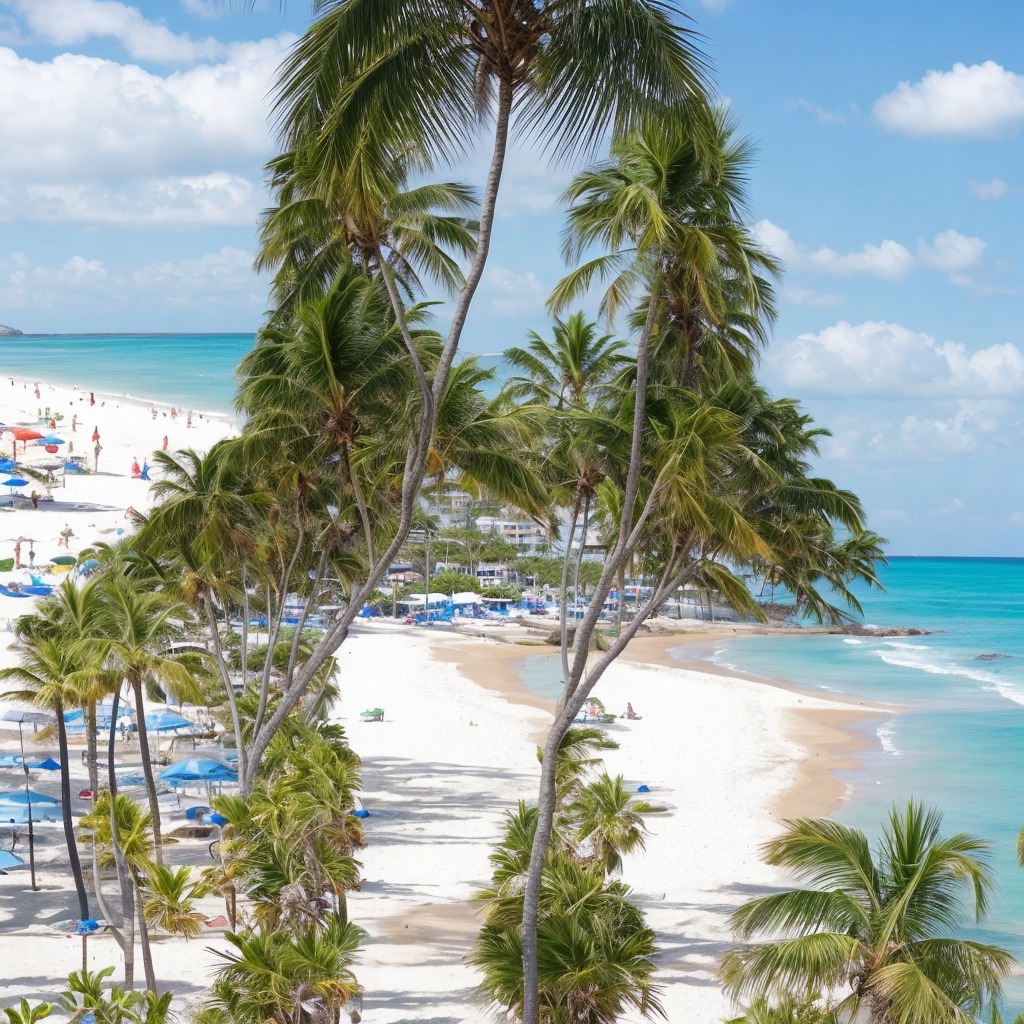}
     \caption{}
     \label{fig:p2}
   \end{subfigure}
    \caption{Images posted to the Facebook that make use religious pareidolia to drive engagement to their posts, both images show images of the face of Jesus appearing in photorealistic imagery. (a) Image posted to the page \textit{Cute Babies} showing two girls standing in water holding hands making the figure of Jesus. (b) Image posted to the page \textit{Interesting stories} of a tropical beach setting with palm trees making out the figure of Jesus. Both images are posted with the caption `Close your eyes 70\% and see magic'.}
    \label{fig:pareidolia}
\end{figure}

Many bot driven accounts make use of the phenomenon of pareidolia of religious iconography to make posts that drive engagement. Figure \ref{fig:pareidolia} shows examples of two posts of photorealistic imagery that have the pareidolia effect of images of Jesus. These are likely done using the method of using a mask pattern to guide a text-to-image diffusion model that was originally developed to make photorealistic images that could act as scannable QR codes \cite{fu2023stylistic}.

Most of the comments on these religious posts are people saying `Amen'. These accounts sharing religious imagery are ultimately taking advantage of people faith and their willingness to performatively engage with these posts and comment on them to demonstrate and reaffirm their faith on social media. These images are created with apparently little regard for their actual content or any sensitivity or moderation of imagery that some might consider blasphemous (such as the infamous `Shrimp Jesus' images (Fig. \ref{fig:shrimp})).

\section{Engagement Hacking as Extrinsic Motivation}

The intention behind the development of these bot accounts is likely not for them to be explicitly creative agents, but to maximise engagement on social media platforms in the most efficient and inexpensive way possible. The speed with with generative deep learning can produce realistic media in a mass produced fashion \cite{smith2023ai} means that this technology has now become the cheapest and fastest way to generate content and drive engagement. Templates for posts and combinations of different concepts can be iterated on extremely quickly, with near instantaneous feedback with likes, comments and shares from other social media users.

\begin{figure}[t]
   % \centering
   \begin{subfigure}[b]{0.225\textwidth}
     \centering
     \includegraphics[width=\textwidth]{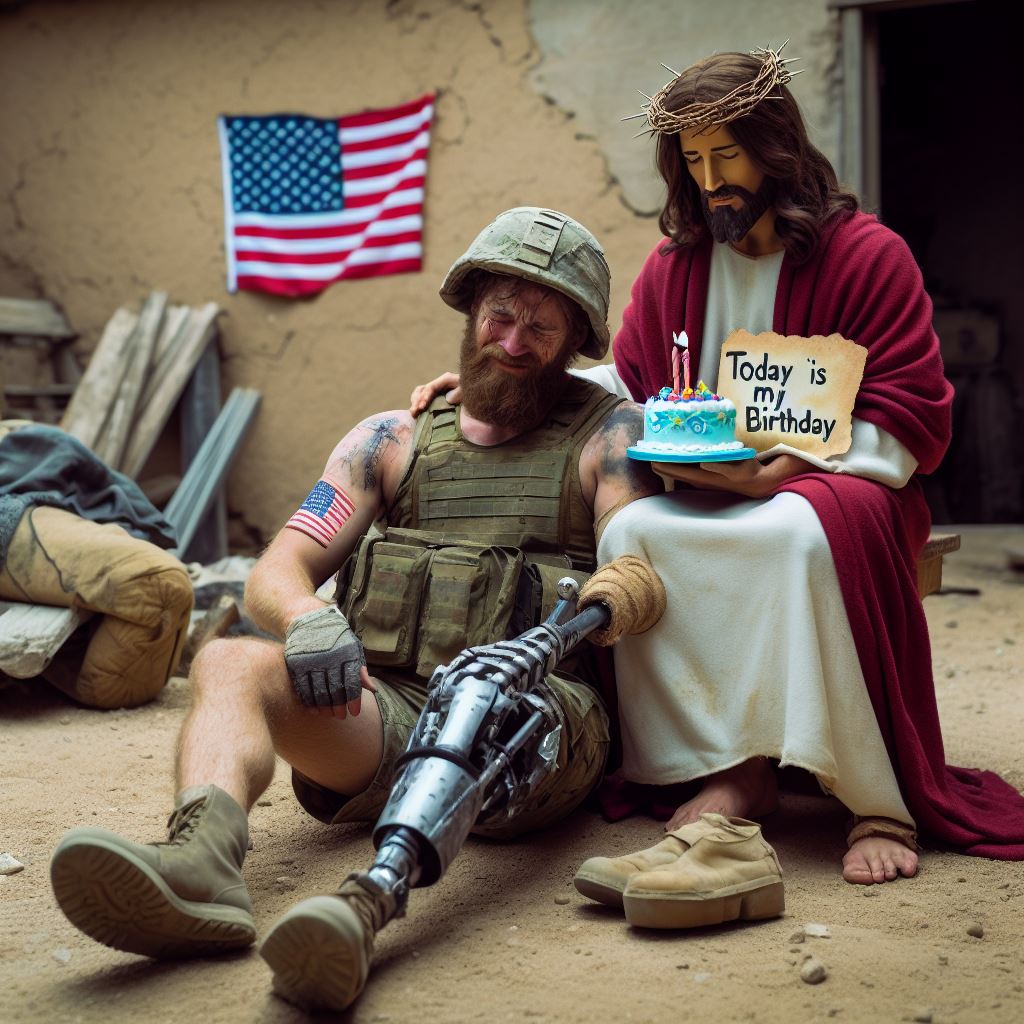}
     \caption{}
     \label{fig:jesus}
   \end{subfigure}
   \hfill
   \begin{subfigure}[b]{0.225\textwidth}
     \centering
     \includegraphics[width=\textwidth]{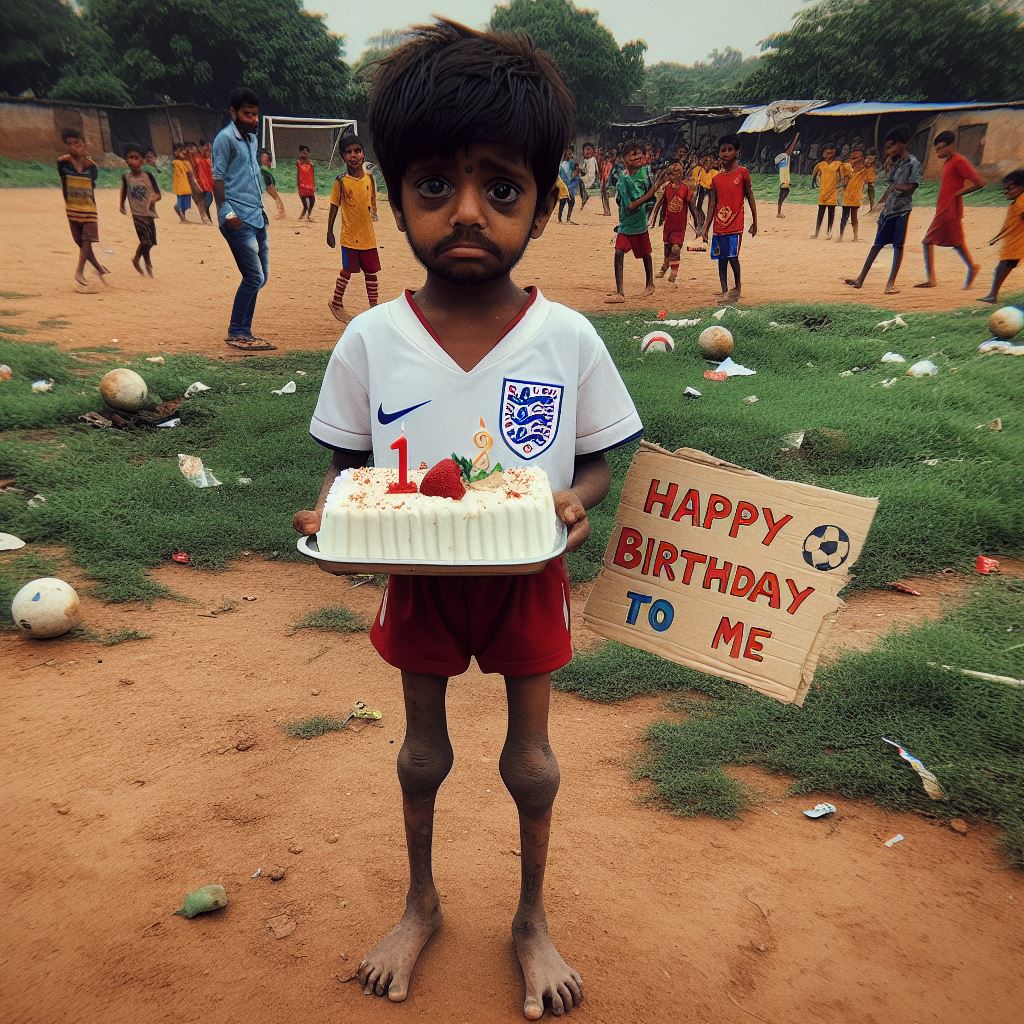}
     \caption{}
     \label{fig:england}
   \end{subfigure}
   \hfill
   \begin{subfigure}[b]{0.225\textwidth}
     \centering
     \includegraphics[width=\textwidth]{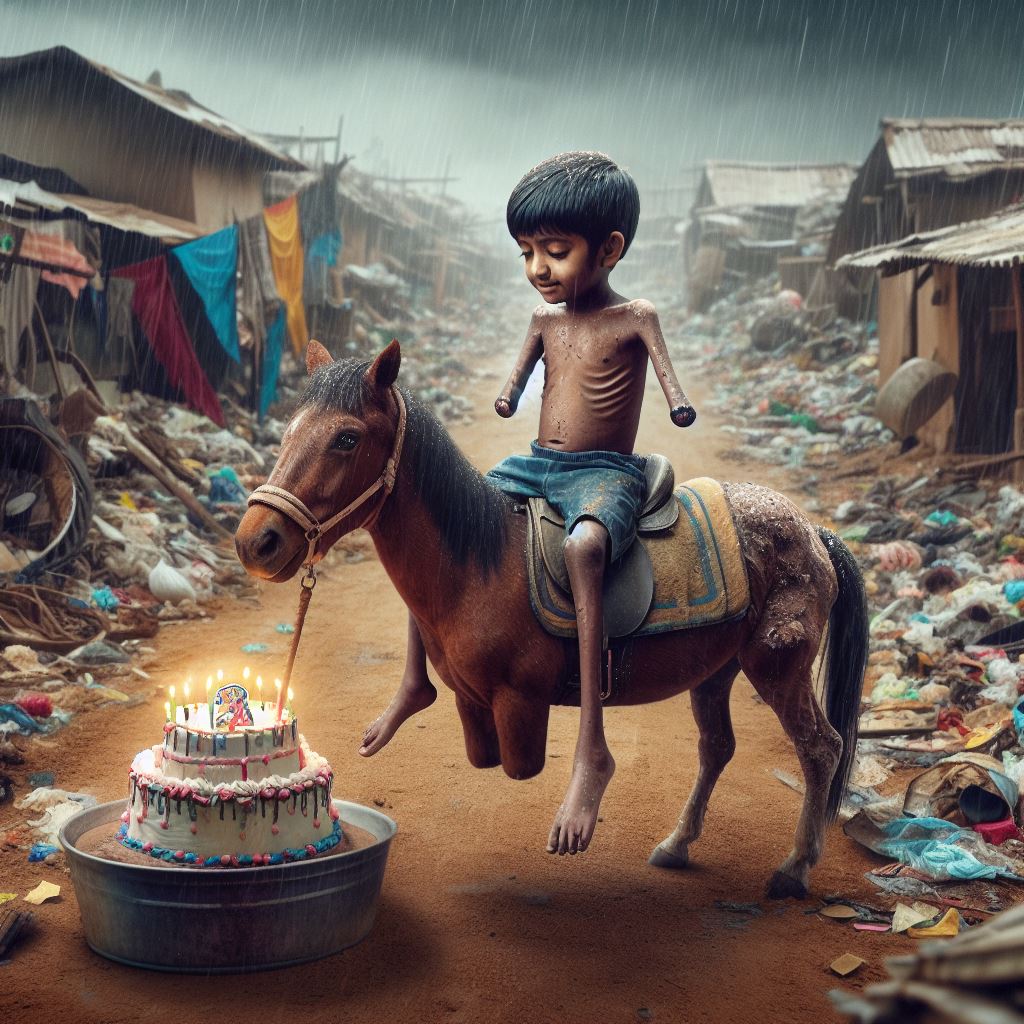}
     \caption{}
     \label{fig:horse}
   \end{subfigure}
   \hfill
   \begin{subfigure}[b]{0.225\textwidth}
     \centering
     \includegraphics[width=\textwidth]{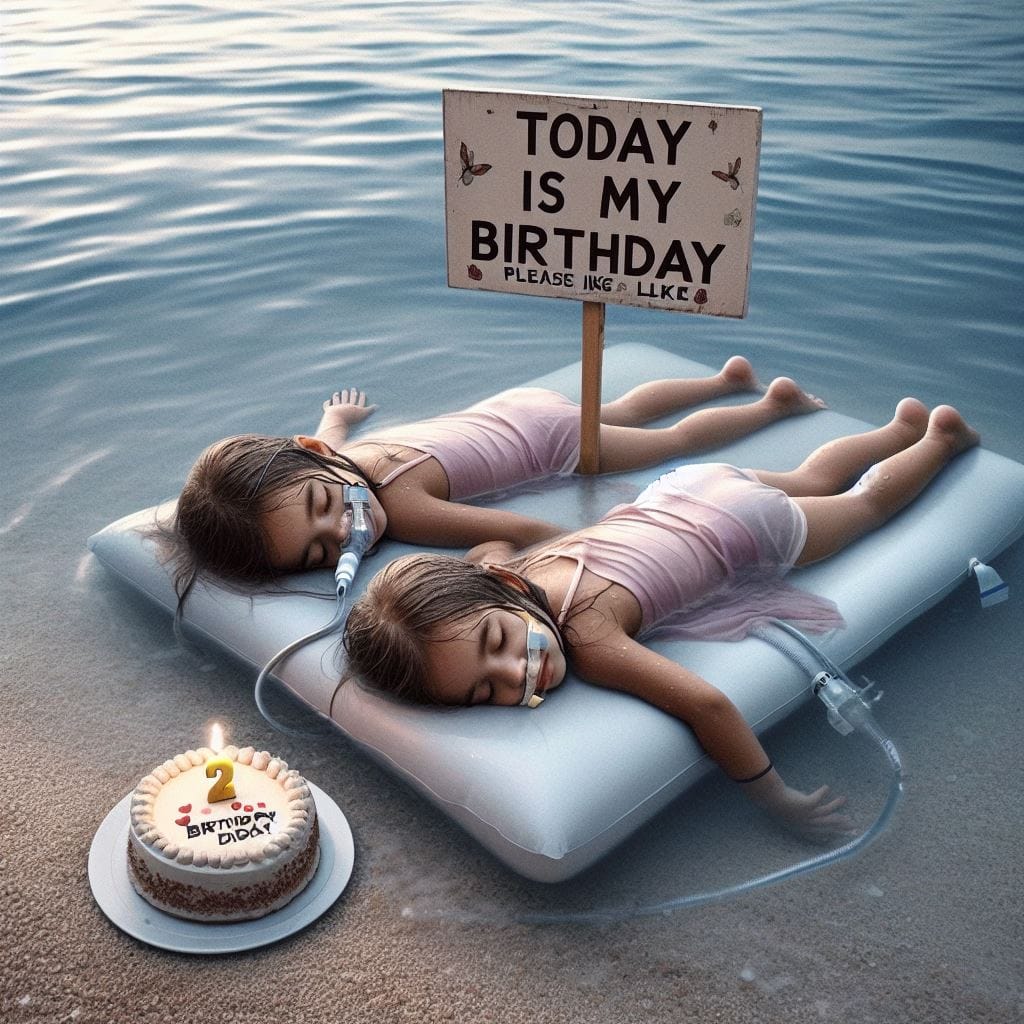}
     \caption{}
     \label{fig:water}
   \end{subfigure}
    \caption{ AI generated images depicting fictitious people on their `birthday' next to birthday cakes posted on Facebook. (a) Image posted to the page \textit{Love dogs Cats Hub} of an injured American soldier with a prosthetic leg next to an image of Jesus holding a birthday cake with a sign saying `Today is my Birthday'. (b) Image posted to the page \textit{Mast} of a small child in a football top holding a birthday cake and a sign saying `Happy birthday to me'. (c) Image posted to the page \textit{Mast} of a small child riding a horse next to a birthday cake, both of which are both missing their lower limbs on their arms and front legs. (d) Image posted to the page \textit{Social help} of two young girls with oxygen masks lying in shallow water with a sign saying `Today is my birthday'. }
    \label{fig:birthday}
\end{figure}

Social media engagement metrics are clearly acting as a means of extrinsic motivation for the agents posting on these accounts in some way. Figure \ref{fig:birthday} depicts images depicting fictitious people in distressing situations who are celebrating their `birthday'. These images are posted with captions like `Happy Birthday To Me, but I haven't received any blessings yet' and are clearly designed in order to deceive users on the platform into liking the posts and commenting birthday wishes for these fictitious people. The images in these posts often contain people with physical ailments and disabilities in order to elicit empathy from users, these ailments include prosthetic limbs (Fig. \ref{fig:jesus}), malnourishment (Fig. \ref{fig:england}), missing limbs (Fig. \ref{fig:horse}), or respiratory problems (Fig. \ref{fig:water}). Alongside this these images will combine nationalistic, religious and sporting symbolism's to further drive engagement to these posts. 

To what extent these accounts are using social media engagement metrics as direct form of extrinsic motivation in a closed autonomous loop, or whether this is mediated by humans managing these bots is difficult to determine. But it is clear that social media engagement and recommendation engine optimisation is the primary goal for these accounts. The quality, coherence and suitability of the content being posted is of a secondary concern and appears to have little human moderation. 

\section{Framing of Authorship}

\begin{figure*}[tbp]
   \centering
   \begin{subfigure}[b]{0.195\textwidth}
     \centering
     \includegraphics[width=\textwidth]{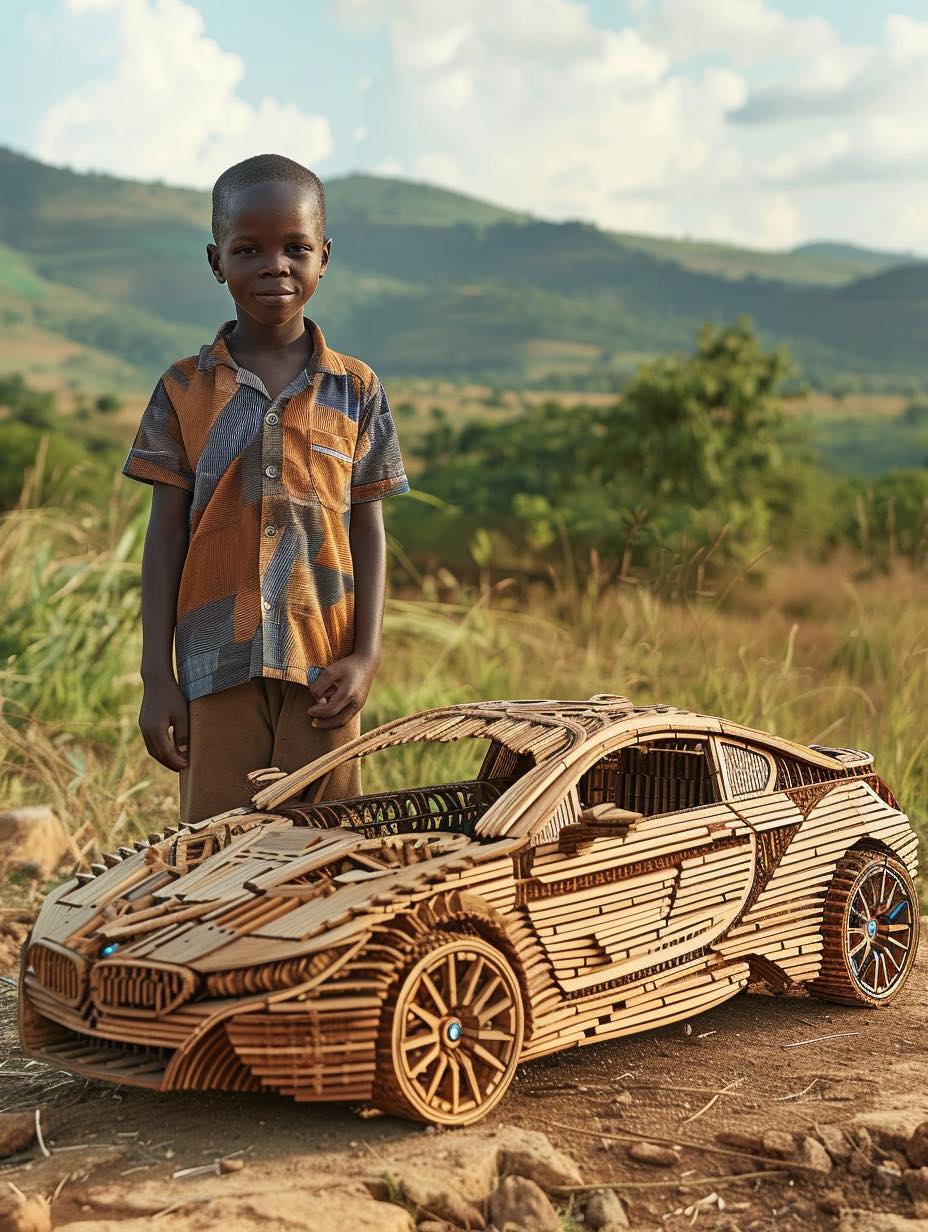}
     \caption{}
     % \caption{I did it myself, I'm waiting for your feedback (Interesting Stories)}
     \label{fig:fruits}
   \end{subfigure}
   \hfill
   \begin{subfigure}[b]{0.195\textwidth}
     \centering
     \includegraphics[width=\textwidth]{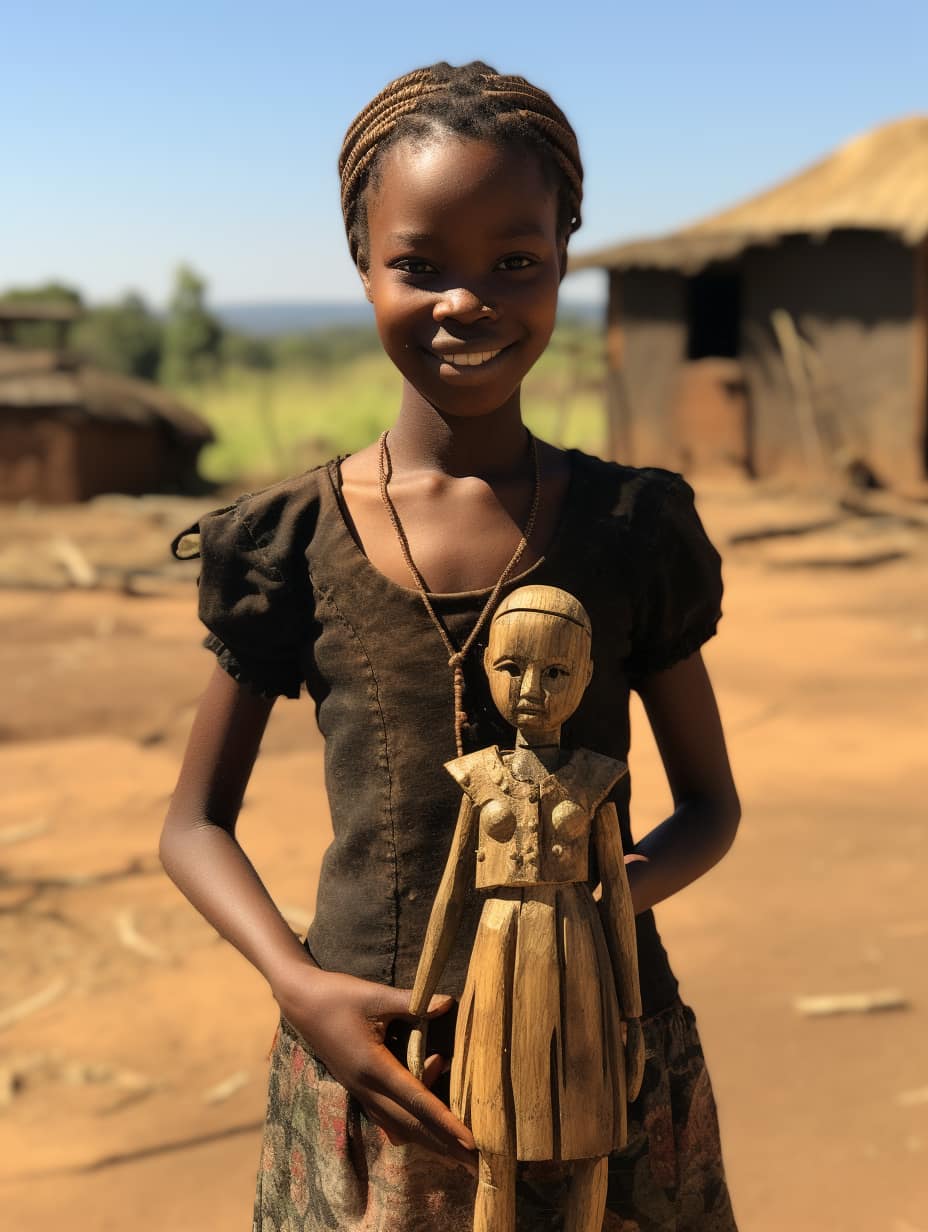}
     \caption{}
     % \caption{I did it myself, I'm waiting for your feedback (Interesting Stories)}
     \label{fig:chaining}
   \end{subfigure}
   \begin{subfigure}[b]{0.195\textwidth}
     \centering
     \includegraphics[width=\textwidth]{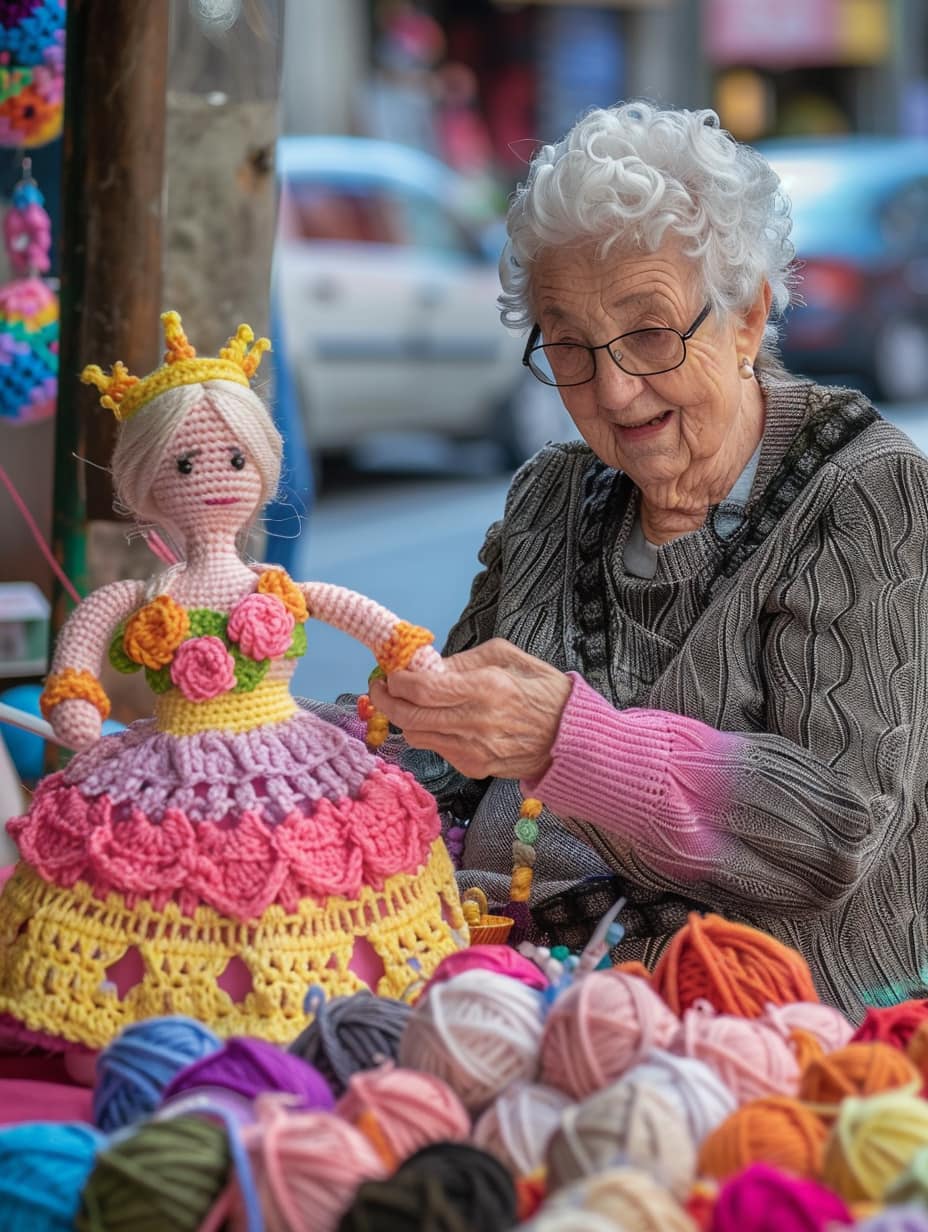}
     \caption{}
     % \caption{I did it myself, I'm waiting for your feedback (Interesting Stories)}
     \label{fig:teratome}
   \end{subfigure}
   \hfill
   \begin{subfigure}[b]{0.195\textwidth}
     \centering
     \includegraphics[width=\textwidth]{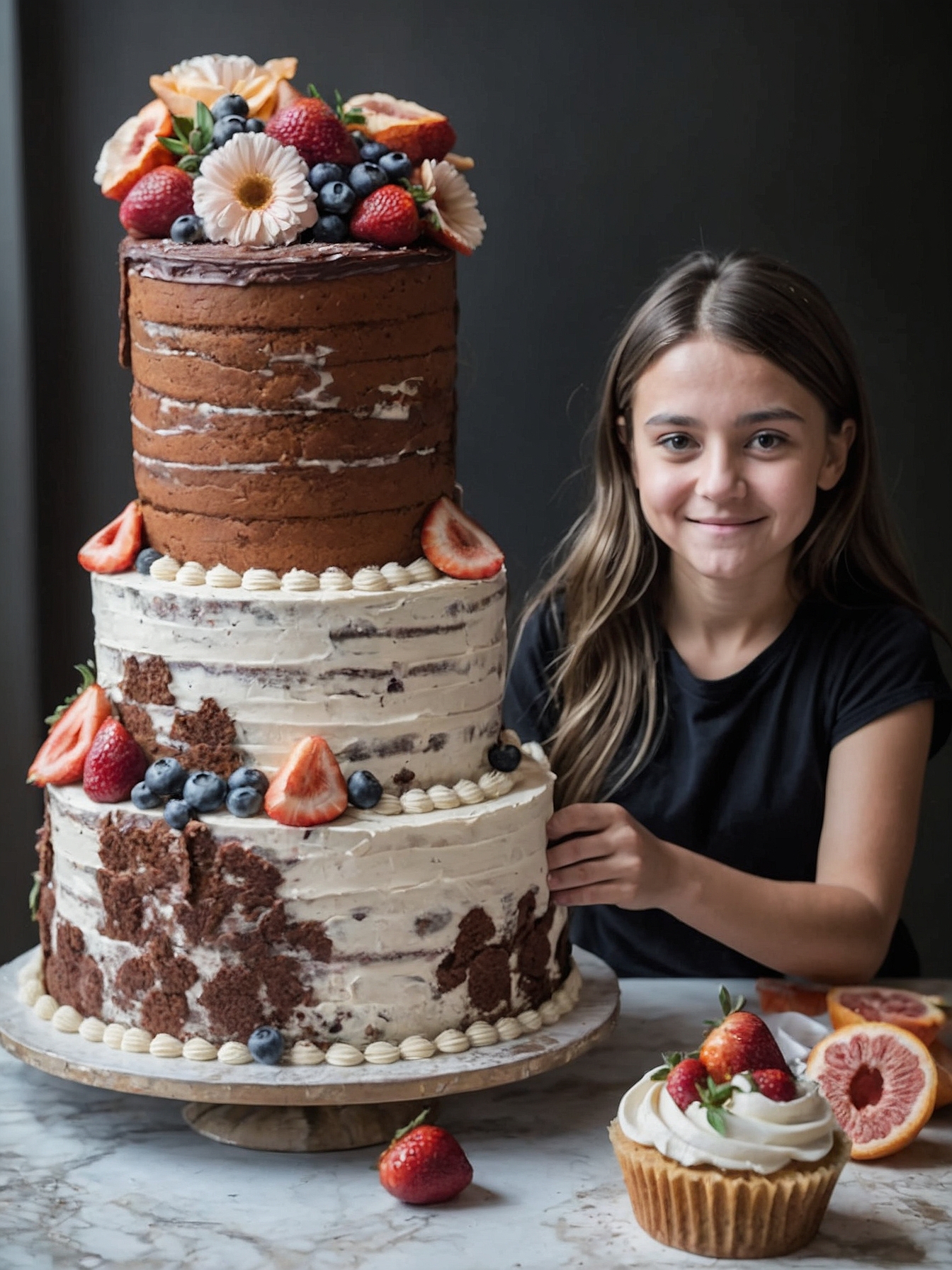}
     \caption{}
     % \caption{did it myself, I'm waiting for your feedback (Interesting Stories)}
     \label{fig:layerswap}
   \end{subfigure}
   \begin{subfigure}[b]{0.195\textwidth}
     \centering
     \includegraphics[width=\textwidth]{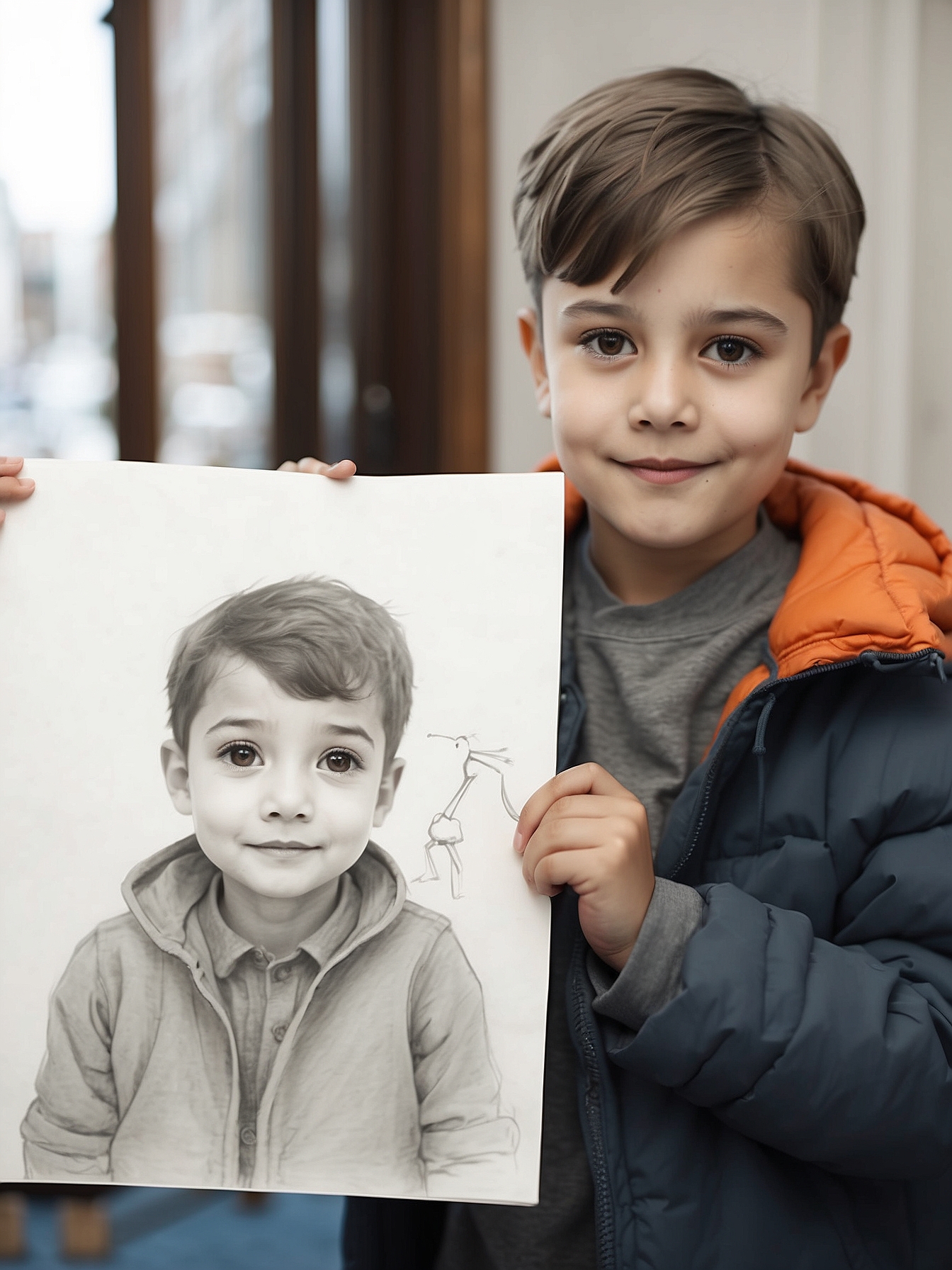}
     \caption{}
     % \caption{did it myself, I'm waiting for your feedback (Interesting Stories)}
     \label{fig:layerswap}
   \end{subfigure}
    \caption{ AI generated images posted on Facebook that include fictitious human creators in the generated images. (a) Image posted on the page \textit{Interesting stories} with the caption `I did it myself, I'm waiting for your feedback'. (b) Image posted on the page \textit{Interesting stories} with the caption: `Made it with my own hands, but no one appreciates this'. (c) Image posted on the page \textit{Interesting stories} with the caption: `This granny wants to be rated for her beautiful crochet work'. (d) Image posted on the page \textit{Life is beautiful} with the caption: `I am taking part in a competition and this is my first work, please appreciate it'. (e) Image posted on the page \textit{Life is beautiful} with the caption: `My son painted his first picture I am very proud of him'. }
    \label{fig:made-with-own-hands}
\end{figure*}

Many of these accounts on Facebook and other social media platforms frame themselves as aggregators of content, which are sharing (or reposting) the work of individuals, in the vein of a  `meme page' or `meme aggregator' \cite{ctuaran2014structure}. The pages themselves are not framing themselves as the creators of the works, but simply the aggregators of content. 

Often times on these pages the work is framed as having a human creator that is present in the generated image. Figure \ref{fig:made-with-own-hands} shows several examples of these images which with the images and associated captions are clear examples of deceptive framing \cite{cook2019framing} being utilised to drive engagement. These images are not of real people but deepfake human avatars that are generated as part of the image, presumably included in the prompt of a text to image generator. 

By including images of the supposed human creators in the images, these bot accounts are seeking to enhance the persuasiveness of human authorship. This is akin to creating works under a pseudonym, but here it is a deepfake human avatar sitting besides the work. The real-world settings is likely designed to enhance supposed authenticity. In most of these posts, there is little other effort is made to frame the works beside a click-bait style caption.

\section{Computational Creativity in the Wild}

With the expense and difficulty in moderating social media platforms, and the difficult in detecting AI generated content, this phenomena is unlikely to disappear anytime soon. Bots are already generating content on social media in at least a semi-autonomous fashion. The computational creativity research community will need to start taking this phenomena seriously. By examining these examples in a critical fashion, this will further our understanding how models of computational creativity are being utilised by nefarious actors to generate spam and and scam content. These bots are utilising creative processes for ulterior motives with little consideration for the suitability of the content being produced. This could be considered a form of dark creativity \cite{cropley2010dark} or malevolent creativity \cite{cropley2013understanding}.

If we are to understand these systems we will need to start analysing them from an external perspective to try to infer how they operate, as the likelihood is that many of the creators of these bot accounts will not be forthcoming about this. Platform restrictions permitting, the behaviour of these bot accounts could be analysed in a systematic way to try and reverse engineer their behaviour, such as prompts used and evolution over time. Leakage of text from prompts into writing on the images may also be a means to investigate the instructions given to these models. In addition, qualitative approaches could also be taken to better understand the behaviour of these bot accounts. 

\section{Conclusion}

This paper suggests that models of computational creativity, whether intentionally or not, are being employed by nefarious actors to maximise engagement on social media in order to generate spam, to underpin content farming operations, and to generate engagement on accounts that can later be used for promoting scams. This is a clear delineation from, and emergence of computational creativity from the traditional worlds of academic research and creative practice, where computationally creative systems have previously been developed and disseminated. The goal of these bots are likely not necessarily to be creative agents per se, but simply a means to an end for some ulterior profit making motive. 

How much of the behaviour of these bots is fully automated, and how much creativity autonomy in turn they have \cite{berns2021automating} is not currently possible to determine. Evaluating the extent of autonomy in these systems requires further investigation. However, the nature of the generations, both in their highly abnormal combinations of concepts, their seemingly regular and mechanical behaviour, and the perceived lack of content moderation of items being posted seem to suggest that these bots are acting in at the very least a semi-autonomous fashion.

As more and more of the internet is being populated by content generated by generative AI and automated bots, moving us closer towards a `dead internet', then the prospect of these agents acting in creative ways, whether intentionally or not should be taken seriously. Studying the behaviour of these bots through the lens of computational creativity will help us understand the impact that these bots are having on both cultural production and the broader media ecology.

\bibliographystyle{iccc}
\bibliography{iccc}

\end{document}